# Room temperature write-read operations in antiferromagnetic memory

Takahiro Moriyama*[1], Noriko Matsuzaki[1], Kab-Jin Kim[1], Ippei Suzuki[2], Tomoyasu Taniyama[2], and Teruo Ono[†1]

[1] *Institute for Chemical Research, Kyoto University, Gokasho, Uji, Kyoto, 611-0011, Japan.*

[2] *Materials and Structures Laboratory, Tokyo Institute of Technology, 4259 Nagatsuta, Midori-ku, Yokohama 226-8503, Japan*

**Abstract**

B2-ordered FeRh has been known to exhibit antiferromagnetic-ferromagnetic (AF-F) phase transitions in the vicinity of room temperature. Manipulation of the Néel order via AF-F phase transition and recent experimental observation of the anisotropic magnetoresistance in antiferromagnetic FeRh has proven that FeRh is a promising candidate for antiferromagnetic memory material. In this work, we demonstrate sequential write and read operations in antiferromagnetic memory resistors made of B2-orderd FeRh thin films by a magnetic field and electric current only. Our demonstration of writing and reading at ambient room temperature opens a realistic pathway towards operational antiferromagnetic memory devices.

* mtaka@scl.kyoto-u.ac.jp
† ono@scl.kyoto-u.ac.jp

Magnetic information storages such as hard disk drives (HDD) and magnetic random access memory (MRAM) have traditionally used a nanometer-scale ferromagnet as a memory bit[1]. Because the magnetic moment in a uniaxially anisotropic ferromagnet has a directional bi-stability, the information can be stored by changing the direction of the magnetic moment and consequently, the memory is nonvolatile. The manipulation of the magnetic moment in ferromagnets can be accomplished by a magnetic field or spin polarized electrons, among other techniques. Their memory states can be read by a stray magnetic field and the magnetoresistance.

An antiferromagnet also has atomic-scale magnetic moments. Neighboring magnetic moments point in opposite directions and therefore, as a whole, an antiferromagnet has no net magnetic moment at its ground state. The absence of a net magnetic moment makes both the manipulation and the reading of the direction of the magnetic moment fundamentally difficult in conventional manners. However, one may notice that a zero stray field and the small susceptibility of antiferromagnets is often advantageous when various field disturbances are a concern. For instance, ultra-high density memory could be possible with antiferromagnetic bits by packing the memory bits to the greatest extent without having to consider the bit-to-bit field interferences[2].

Recently, the manipulation of the direction of antiferromagnetic moments via AF-F phase transition of FeRh and observation of the antiferromagnetic anisotropic magnetoresistance (AMR) has demonstrated that antiferromagnets have a significant potential as a novel memory resistor[3]. B2-orderd FeRh is antiferromagnetic at room temperature and undergoes the transition to the ferromagnetic phase at approximately 380 K[4,5,6]. At a temperature above the phase transition temperature, FeRh becomes ferromagnetic and the magnetic moment can be controlled by an external field. By maintaining the external field and cooling below the transition temperature, the magnetic moments align perpendicularly to the external field. With this field cooling process, the Néel order, or the antiferromagnetic state, can be written by the direction of the external field. Marti *et al.*[3] demonstrated the resistance change associated with this field-cooling process with different directions of the applied field, demonstrating the write and read of an antiferromagnetic state. However, the external temperature control of the AF-F phase transition is hardly compatible with possible memory applications. In fact, the AF-F phase transition can be driven by other factors such as strain[7,8,9], spin polarized current[10], a strong external magnetic field[6,11], and current-induced Joule heating[12]. Among these, current-induced AF-F phase transition is one of the most accessible techniques for realistic memory applications[12].

In this work, we demonstrate sequential write-read operations of antiferromagnetic memory resistors by electric current and external field only. Upon writing, the AF-F phase transition is induced by Joule heating by the electric current flowing through an FeRh wire. We adapted a resistance "ratio" measurement technique to capture the small resistance change due to the antiferromagnetic AMR. All the operations were performed at an ambient room temperature; no elevating external temperature was required to realize the write-read operations.

We prepared a B2-ordered epitaxial 42 nm FeRh thin film capped with 10 nm $SiO_2$ on a 500 μm thick MgO(001) substrate. For the detail growth condition and the quality of the FeRh, one should reference to Ref. 13. The film was patterned into the crossed-wire structure shown in Fig. 1(a). A d.c. current $I_{dc}$ flows in the portion of the crossed wire indicated by the dotted square in Fig. 1(a). The voltage *V* on the horizontal branch of the wire and the reference voltage $V_{ref}$ on the vertical branch are simultaneously measured as the ratio $V/V_{ref}$ by a voltmeter capable of ratio measurement (We used a Keithley 2182 nanovoltmeter). The voltage ratio is then translated to the resistance ratio $R/R_{ref}$ by $I_{dc}$. This technique allows us to sensitively differentiate a small resistance change of interest from unwanted noises by cancelling the common mode noises in the two branches. We found this technique was particularly effective for an FeRh wire system having a strong resistance drift due to the ambient temperature fluctuation. We considered two configurations of the Néel order, which can be defined by the direction of the field upon the field-cooling process as indicated in Fig. 1(b) and (c). The $R/R_{ref}$ of the two configurations differs by $\delta = R_\parallel/R_\perp - R_\perp/R_\parallel$, where $R_\parallel$ and $R_\perp$ is the resistance when the Néel order is parallel and perpendicular to the current flow, respectively. The actual resistance change due to the AMR in the horizontal branch $\Delta R = R_\parallel - R_\perp$ is calculated by $\Delta R = R_{ref}\delta/2$. We define the magnetoresistance as $MR = \Delta R/R_\perp$ in this report. We prepared and investigated various wire widths in the range 1~5 μm. We found that the fundamental mechanisms of the write-read are independent of the wire width; hence, we focus on the typical results obtained from 1 μm-wide wires in the following discussion.

Current-induced AF-F phase transition can be seen in the *I-R* characteristics of the two wire branches as shown in Fig. 2. Both branches indicate significant resistance drops near $I = \pm 8$ mA, which are characterized by the AF-F phase transition. The resistance in the ferromagnetic state is lower than that in the antiferromagnetic state[5,14,15,16]. It is evident that the FeRh wire is in the antiferromagnetic state with $|I|$ less than 7 mA and is in the F state with $|I|$ greater than 12 mA. Because the AF-F phase transition in FeRh is classified as a first order phase transition[17], it generally exhibits a hysteresis behavior that can also be seen in the *I-R* characteristic. The multistep resistance change during the phase transition may originate from the uneven width at the intersection of the wire, as the critical current is dependent on the wire width[12]. Our previous study on a single straight FeRh wire indeed showed smoother resistance change[12].

Figure 3 illustrates the procedure of the write and read cycle performed in this work. To begin, with no external field, we apply current $I_{Hi}$. The FeRh becomes ferromagnetic if $I_{Hi}$ is sufficiently large to invoke the AF-F phase transition. While maintaining $I_{Hi}$, we apply an external field in either the *x* or *y* direction denoted as $H_x$ or $H_y$. The coordinate system is defined in Fig. 1(a). While maintaining the external field, we reduce the current to $I_{Lo} = 0.5$ mA for reading the resistance. The FeRh now becomes antiferromagnetic. We turn off the external field after a few seconds and wait for the heat to dissipate for approximately 30 second before reading the resistance. The resistance is read by the ratio measurement mentioned above and the measurement is repeated 50 times in each write-read cycle. The direction of the external field is interchanged in every write-read cycle.

Figures 4 present the resistance change measured with the sequential write-read operations. The measurements begin by writing with $I_{Hi} = 5\sim13$ mA and $H_y = 3$ kOe followed by 50 resistance readings. Then, another write operation with $I_{Hi} = 5\sim13$ mA and $H_x = 3$ kOe is performed followed by 50 resistance readings. When $I_{Hi} = 13$ mA, which is sufficiently large to bring the FeRh wire into the ferromagnetic phase, the resistance reads a high state after the write operation with $H_y = 3$ kOe and reads a low state after the write operation with $H_x = 3$ kOe as indicated in Fig. 4(a). We observed a resistance change as much as 45 mΩ between the two operations, which translates to 0.011% magnetoresistance. When $I_{Hi} = 8$ mA, which is approximately the boundary of the phase transition, the resistance change continues to be observed upon the write operation, however, it is smaller than the case with $I_{Hi}$ = 13 mA. This indicates that the ferromagnetic phase is partially formed with $I_{Hi}$ = 8 mA and the Néel order is altered only in the portion undergoing the phase transition. This is consistent with the characteristics of a first order phase transition where the antiferromagnetic and ferromagnetic phase can coexist during the transition. We found that the small resistance change continues to exist even with $I_{Hi}$ = 5 mA with which the AF-F phase transition is not supposed to be induced. This may suggest that there is a ferromagnetic phase at room temperature due to an imperfect antiferromagnetic ordering of the FeRh film, which exhibits the ferromagnetic AMR[3]. To confirm that the above resistance changes are because of the magnetoresistance, we rotated the coordinate axes by $\theta$ = 45º as depicted in Fig. 1(a) and conducted the write-read operations with $I_{Hi}$ = 13 mA. In this configuration, the Néel order always makes 45º with respect to the current flowing direction and therefore no resistance change between $H_x$ and $H_y$ write operations should be observed. The results clearly indicate that the resistance change is minimized as plotted in Fig. 4(d).

Figure 5(a) summarizes the observed MR in the write-read cycles as a function of the magnitude of $I_{Hi}$. The MR saturates at 0.011% with $I_{Hi}$ > 8.5 mA or 2.0 x $10^{11}$ A/m$^2$, which is considered as the critical current for writing in this particular FeRh wire. It is interesting to note that the MR saturates at a current before the AF-F phase transition completes (compare Fig. 2 and Fig. 5(a)). On the other hand, the onset current $I_{Hi}$ of the MR increase corresponds to the onset current of the AF-F phase transition. This indicates that a complete transformation into the ferromagnetic phase is not necessary to manipulate the final antiferromagnetic state. That is, the partial transformation of the ferromagnetic phase is sufficient to control the remainder of the antiferromagnetic phase by the external field, implying the interphase magnetic coupling between antiferromagnetic and ferromagnetic phases may have an important role. As mentioned previously, the residual MR with $I_{Hi} = 0$ mA can be attributed to the residual low temperature ferromagnetic phase.

We also investigated the $H_{x,y}$ dependence of the MR with $I_{Hi} = 13$ mA and 0 mA as illustrated in Fig. 5(b). The MR slowly saturates toward $H_{x,y} = 3$ kOe with $I_{Hi} = 13$ mA, suggesting that one should be cautious of the intensity of the applied field on the write operation. The MR of 0.003% with $I_{Hi} = 0$ mA is attributed to the ferromagnetic AMR in the residual ferromagnetic phase. By subtracting this ferromagnetic AMR contribution, the actual antiferromagnetic AMR is ~0.008%, which is quite small compared to that reported by Marti et al.[3] of ~0.4% antiferromagnetic AMR in the Rh-rich FeRh film. The richer the Rh composition, the larger the antiferromagnetic AMR predicted in Ref. 3 and Ref. 18. It is reasonable that our $Fe_{50}Rh_{50}$ demonstrates a small MR.

In a naïve picture, the Néel order should be formed by spin flopping with respect to the applied field direction during the ferromagnetic to antiferromagnetic phase transition. Therefore, the $H_{x,y}$ required to set the Néel order should be equivalent to the saturation field of the F state. Compared to the AMR curve in the high temperature ferromagnetic phase (indicated in the inset of Fig. 5(b)), which saturates near 1 kOe, the slower saturation of the MR with respect to $H_{x,y}$ is rather counterintuitive. This suggests that one should consider more complex interplay during the field cooling among the temperature dependent exchange energies in both antiferromagnetic and ferromagnetic states, Zeeman energy, and anisotropy energies as well as the interphase magnetic interactions.

Finally, we demonstrate the stability of the antiferromagnetic memory against external field disturbances. We first confirm the magnetoresistance in the write-read cycles by performing the two write operations. With the established "high" resistance state, the disturbing fields in the $x$ and $y$ directions, respectively denoted as $H_{dist,x}$ and $H_{dist,y}$, are consecutively applied. The test results are presented in Fig. 6. Although the MR indicates minor variability when $H_{dist} > 2$kOe, the MR does not reduce to the lowest state. Moreover, the extent of the minor MR variability with $H_{dist} > 2$kOe is comparable to that observed in Fig. 5(b) with $I_{Hi} = 0$ mA, indicating that the observed MR reduction with $H_{dist} > 2$kOe could be a consequence of the residual ferromagnetic phase. The antiferromagnetic Néel order is fundamentally unchanged with our tested disturbing field range.

In summary, we demonstrated the room temperature write-read operation of antiferromagnetic memory made of B2-ordered FeRh. We successfully wrote the memory resistor using electric current and an external field only and read the resistor using the antiferromagnetic anisotropic magnetoresistance. The writing mechanism relied on the AF-F phase transition induced by Joule heating of the resistor. We demonstrated that the writing current was $2.0 \times 10^{11}$ A/m$^2$ in the 1 μm-wide wire. In principle, the writing current can be reduced by further narrowing the wire width[12]. Our results open a realistic pathway towards operational antiferromagnetic memory devices.

**Acknowledgements**

This work was supported by JSPS KAKENHI Grant Numbers 26870300, 15H05702, 26289229.

**Figure captions**

Figure 1 (a) Schematic of the measurement setup. (b) Portion of the crossed wire indicated by the red dotted square in (a). When the field direction during the field cool $H_{FC}$ is in $y$ direction, the Néel order is set in $x$ direction. (c) When $H_{FC}$ is in $x$ direction, the Néel order is set in $y$ direction.

Figure 2 Current-resistance characteristic of the FeRh wire. The colored hatchings indicate the antiferromagnetic phase (AF) and ferromagnetic phase (F).

Figure 3 Procedure of write and read operations. The red hatching indicates the write operation and the blue hatching indicates the read.

Figure 4 Resistance change as a function of the number of readings. The resistance reading of 50 times is following the writing operation with $H_{x,y} = 3$ kOe. $I_{HI}$ and $\theta$ are varied as (a) $I_{HI} = 13$ mA and $\theta = 0°$, (b) $I_{HI} = 8$ mA and $\theta = 0°$, (c) $I_{HI} = 5$ mA and $\theta = 0°$, (d) $I_{HI} = 13$ mA and $\theta = 45°$.

Figure 5 (a) Magnetoresistance as a function of $I_{HI}$. (b) Magnetoresistance as a function of $H_{x,y}$. The inset indicates the magnetoresistance curve of the ferromagnetic phase with the external field swept in the $y$ direction. The measurement was performed at 13 mA where the ferromagnetic phase was stable.

Figure 6 Stability test results plotting the magnetoresistance as a function of the number of readings. The disturbing fields $H_{dist,x}$ and $H_{dist,y}$ are consecutively applied following the write processes.

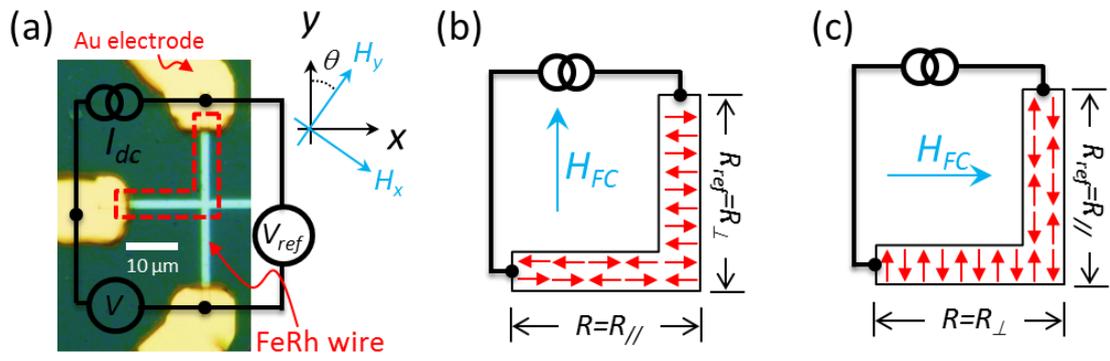

Figure 1 Moriyama *et al*.

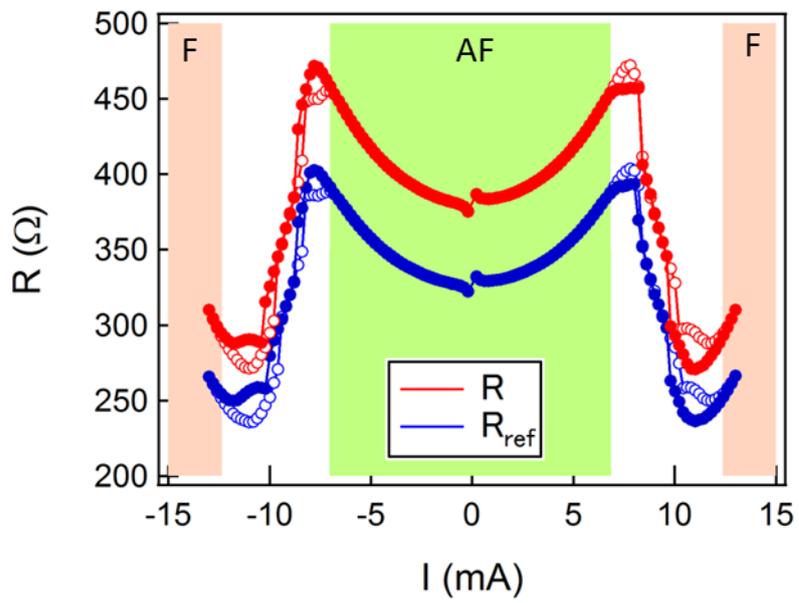

Figure 2 Moriyama *et al*.

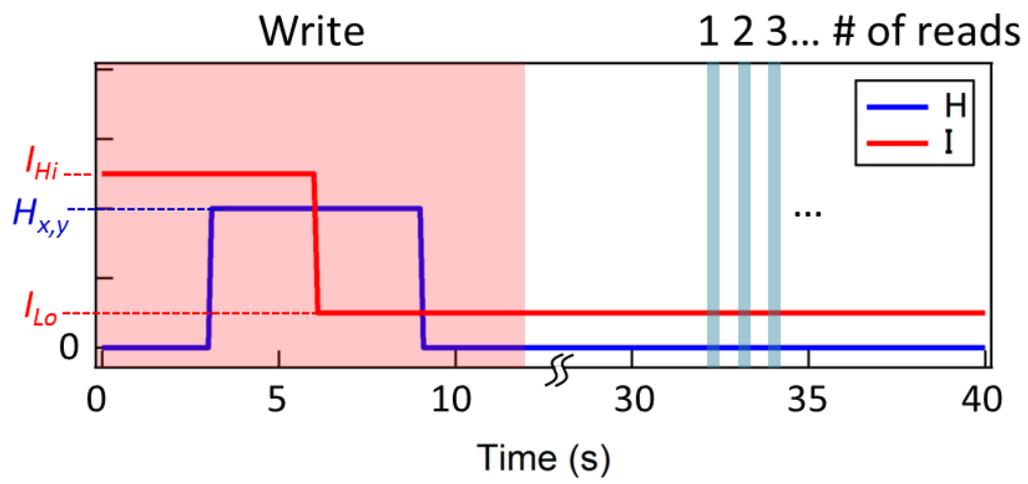

Figure 3 Moriyama *et al*.

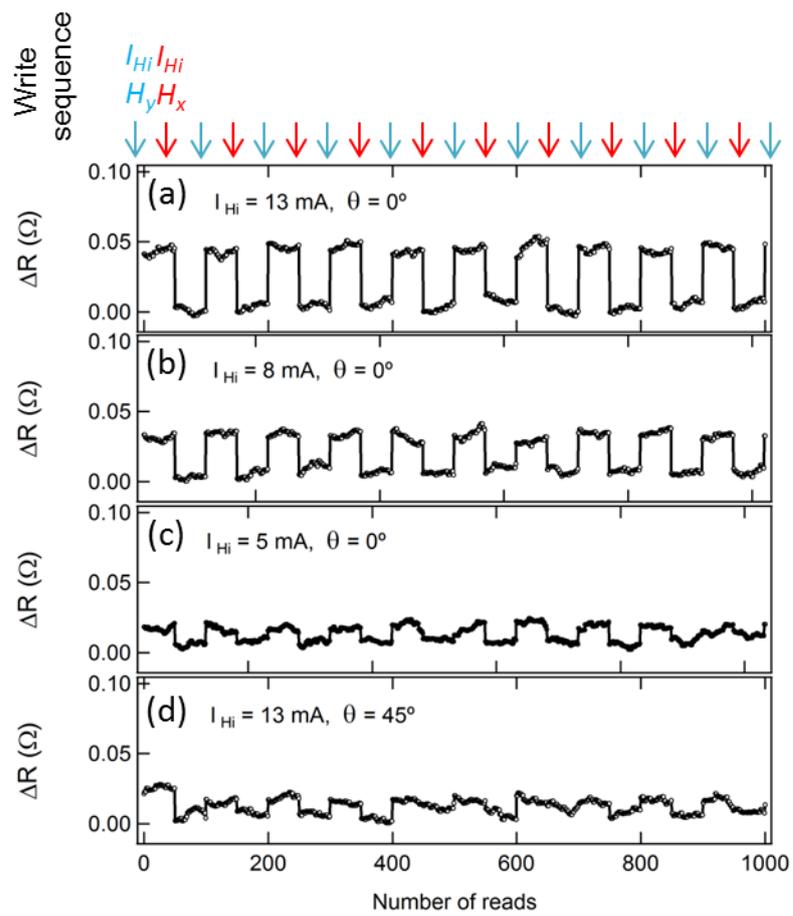

Figure 4 Moriyama *et al*.

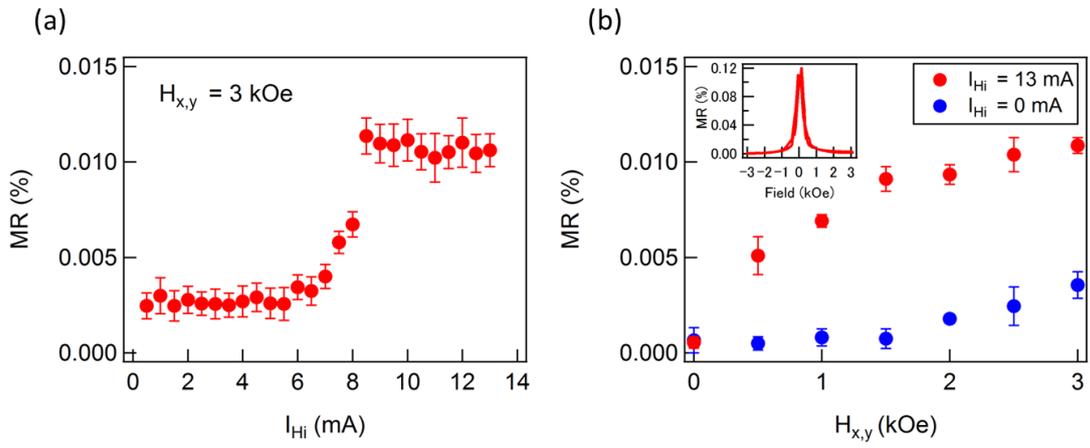

Figure 5 Moriyama *et al*.

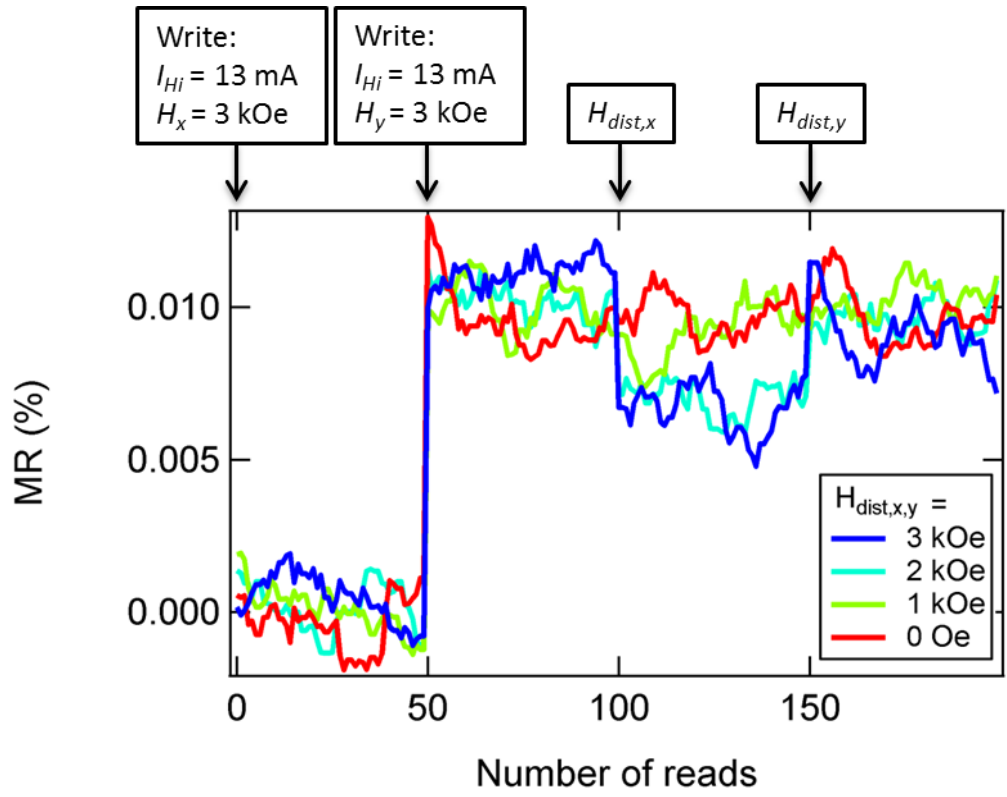

Figure 6 Moriyama *et al*.